\documentclass[aps]{revtex4-1}
\usepackage{graphicx}
\usepackage{amsmath}
\usepackage{amssymb}
\usepackage{tikz}
\usepackage{float}
\usepackage{sidecap}
\usepackage{multirow}
\usepackage[caption = false]{subfig}
\usepackage{tabularx,booktabs}
\newcolumntype{Y}{>{\centering\arraybackslash}X}
\usepackage[margin=1in]{geometry} 
\usetikzlibrary{fit,calc,positioning,decorations.pathreplacing,matrix}

\begin{document}
\title{\bf Low-energy constants of heavy meson effective theory in lattice QCD } 
\author{Mohammad H. Alhakami}
\affiliation{Nuclear Science Research Institute,
KACST, P.O. Box 6086, Riyadh 11442, Saudi Arabia}
\date{\today}
\begin{abstract}
We consider effective theory treatment for the lowest-lying $S$- and $P$-wave states of charmed mesons.
In our analysis, quantum corrections and contributions from leading chiral and heavy quark symmetry breakings are taken into account.
The heavy meson mass expressions have abundance parameters, low-energy constants, in comparison to the measured charmed mesons masses. 
The experimental and lattice QCD data on charmed meson spectroscopy are used to extract, 
for the first time, the numerical values of the full set of
low-energy constants of the effective chiral Lagrangian. 
Our results on these parameters 
can be used for applications 
on other properties of heavy-light meson systems.
\end{abstract}
\pacs{Valid PACS appear here}
\maketitle
\onecolumngrid
\section{Introduction}
The properties of heavy-light meson systems can be well described using heavy meson chiral perturbation theory (HM$\chi$PT). 
This approach, which is formulated by combining chiral perturbation theory ($\chi$PT) and heavy quark effective theory (HQET), 
can be used  in a systematic way
to calculate the corrections from  
chiral and heavy quark symmetry
breakings (see, e.g., Refs. \cite{4,cas97,FalkLuke,PCh,wise}). 
Thomas Mehen and Roxanne Springer in Ref. \cite{ms05} used this theory to study the masses of the lowest-lying odd- and even-parity charmed mesons.
In their analysis, the contributions due to finite masses of light and heavy quarks and one-loop chiral corrections are taken into account.
The theory at this, third, order has a large number of unknown low-energy constants (LECs) in comparison to the charmed meson spectrum, and hence
a unique fit for them using nonlinear fitting is impossible as concluded in Refs. \cite{ms05, absu07}.

The work of Mehen and Springer 
is reconsidered in our paper \cite{Alhakami}.
There, we employed a different approach to get a unique fit for these unknown LECs. 
It is based on 
reducing their number in fit, which is simply done by grouping them into certain linear combinations that equal the number of charmed meson masses,
and evaluating the one-loop corrections using physical masses, 
which, unlike previous approaches, ensures that the imaginary parts of loop functions are consistent with the experimental
widths of the charmed mesons. 
By using physical masses in loops, the fit becomes linear, and 
LECs of the effective Lagrangian, which appear in linear combinations,
are uniquely determined using the lowest odd- and even-parity charmed spectrum.
The fitted parameters from charmed mesons are then used in Ref. \cite{Alhakami}  
to predict the spectrum of analog bottom mesons.

It is pointed out in our previous work 
that 
to separate the combinations of the LECs into pieces that respect and
break chiral symmetry, lattice QCD (LQCD) information on charmed mesons ground and excited states with different quark masses are required. 
The recent lattice calculations on the charmed meson spectroscopy undertaken by Cichy \textit{et al.} in Ref. \cite{lattice} provide enough 
information to perform further separations of LECs. Our purpose here is to use the experimental and these lattice data on charmed meson masses
to extract, for the first time, the unique numerical values of LECs of the effective Lagrangian used in Refs. \cite{ms05,Alhakami,absu07}.

The work undertaken in the present paper is complementary to our previous approach in Ref. \cite{Alhakami}
and organized as follows. In Sec. II we briefly review the mass expressions for the lowest-lying $S$- and $P$-wave states of charmed mesons
that are derived within the framework of HM$\chi$PT. 
We demonstrate how terms in these mass expansions link to experimental measurements on such systems. 
We describe in Sec. III the approach we have employed to extract the unique numerical values for the full set of LECs
of the chiral Lagrangian. It relies on making constraints on certain combinations of LECs using the charmed meson spectrum
and then utilizing lattice data on charmed mesons ground and excited states to disentangle chirally symmetric LECs from chiral breaking terms.  
After presenting the results on LECs, 
we draw our conclusion.  
\section{Low-energy constants in HM$\chi$PT}
Before proceeding, let us first present the mass formula for odd- and even-parity charmed mesons 
that are used 
in Refs. \cite{ms05,Alhakami,absu07}.
In a compact form,  
the residual charmed meson mass \cite{rm} is 
\begin{equation}\label{eq1}
 m_{A^{(*)}_q}=\delta_A+a_A m_q+\sigma_A \overline{m}+\frac{d^{(*)}}{4}(\Delta_A+\Delta^{(a)}_A m_q+\Delta^{(\sigma)}_A \overline{m})+\Sigma_{A^{(*)}_{q}},
\end{equation}
where $A=H,S$ denote the odd- and even-parity charmed meson states, respectively. 
In the heavy quark limit, the odd-parity states, i.e., pseudoscalar mesons $J^P=0^-$ ($D^0,D^+,D^+_s$) and  vector mesons $J^P=1^-$ ($D^{*0},D^{*+},D^{*+}_s$),
form members of the $\frac{1}{2}^-$-ground-state doublet, and 
the even-parity states, i.e.,
scalar mesons $J^P=0^+$ ($D^{*0}_0,D^{*+}_0,D^{*}_{0s}$) and axial vector mesons $J^P=1^+$ ($D^{\prime 0}_1,D^{\prime 1}_1,D^{\prime 0}_{1s}$),
form members of the  $\frac{1}{2}^+$-excited-state doublet.
The asterisk represents the spin-1 meson  in both sectors 
and
the subscript $q$ refers to the flavor of light-quarks. 
The values of the factor $d^{(*)}$ are $1$ for the spin-1 particles ($d^*=1$) and 
$-3$ for the spin-0 particles ($d=-3$). The quantities $m_q$ and 
$\overline{m}$ define as $m_q=(m_u,m_d,m_s)$ and  $\overline{m}=m_u+m_d+m_s$, respectively.
In the isospin limit, $m_u=m_d=m_n$, and hence $m_q=(m_n,m_n,m_s)$ and $\overline{m}=2m_n+m_s$,
where the subscripts $n$ denote nonstrange light quark flavor. We work in the isospin limit.
The parameter $\delta_A$ represents the residual masses of charmed mesons in sector $A$. The
operator $\Delta_A$ gives rise to the hyperfine splittings
at leading order in the chiral expansion. The quantities $a_A$ ($\Delta^{(a)}_A$) and $\sigma_A$ ($\Delta^{(\sigma)}_A$) are dimensionless constants, 
and $\Sigma_{A^{(*)}_{q}}$ 
refers to the one-loop corrections.
According to the power counting rules employed in Refs. \cite{ms05,Alhakami}, these coefficients 
scale as $\delta_A \sim \Delta_A \sim \Delta_A^{(a)} \sim \Delta_A^{(\sigma)}  \sim Q$,
$m_q \sim \overline{m} \sim Q^2$, and $\Sigma_A \sim Q^3$, where $Q$ generically denotes the
low-energy scales in the theory, i.e., masses and momenta of the Goldstone bosons
and splittings between the four lowest states of the charmed mesons introduced above. 

The one-loop corrections can be obtained by adding all one-loop graphs that are allowed by spin-parity quantum numbers.
Their explicit expressions can be found in the Appendices of Refs. \cite{ms05,Alhakami}.
There are three coupling constants $g$, $g^\prime$, $h$  entering the one-loop contributions.
The coupling $g$ ($g^\prime$)
measures the strength of transitions within states that belong to
the $\frac{1}{2}^-$ ($\frac{1}{2}^+$) doublet which are represented by
the chiral function 
\begin{equation}\label{K1}
\begin{split}
K_1(\omega, m_i,\mu)&=\frac{1}{16 \pi^2}\left[(-2\omega^3+3m_i^2\omega)\mathrm{ln}\left(\frac{m_i^2}{\mu^2}\right)-4(\omega^2- m_i^2)F(\omega,m_i)+\frac{16}{3}\omega^3-7\omega\, m_i^2\right],\\[2ex]
\end{split}
\end{equation}
which is defined in the $\mathrm{\overline{MS}}$ scheme \cite{Alhakami}. The renormalization scale is given by $\mu$.
The arguments 
$m_i$ and $\omega$ are the mass of the Goldstone boson and mass difference between external and internal heavy meson
states. The function $F(\omega,m_i)$ is given by \cite{Ffunction}
\begin{equation}\nonumber
 F(\omega,m_i)=\left\{ \begin{array}{c c}-\sqrt{m_i^2-\omega^2} \cos^{-1}(\frac{\omega}{m_i}), & \mbox{$m_i^2>\omega^2,$} \\[2ex]
                                 \sqrt{\omega^2-m_i^2}[i \pi-\cosh^{-1}(-\frac{\omega}{m_i})], & \mbox{$\omega<-m_i,$}\\[2ex]
 \sqrt{\omega^2-m_i^2}\cosh^{-1}(\frac{\omega}{m_i}), & \mbox{$\omega>m_i.$}
                                \end{array} \right.
                                  \end{equation}
The transitions between states that belong to different doublets are measured by the coupling strength $h$
and represented by
\begin{equation}\label{K2}
\begin{split}
K_2(\omega, m_i,\mu)&=\frac{1}{16 \pi^2} \left[ (-2\omega^3+ m_i^2\omega)\mathrm{ln}\left(\frac{m_i^2}{\mu^2}\right)-4\omega^2 F(\omega,m_i)+4\omega^3- \omega\, m_i^2\right],
\end{split}
\end{equation}
in the $\mathrm{\overline{MS}}$ scheme \cite{Alhakami}. 

Let us now briefly show how terms in the above mass expansion,  Eq.~\eqref{eq1}, are linked to the experimental measurements on the heavy-light
meson systems. 
Terms with the coefficients $\delta_A$ and $\sigma_A$ give the same contributions to heavy meson masses.
The $SU(3)$ mass splitting between strange and nonstrange heavy charmed mesons is due to 
$a_A$. Other terms which contain $\Delta_A$, $\Delta_A^{(\sigma)}$, and $\Delta_A^{(a)}$ contribute to chirally symmetric,
chiral symmetry breaking, and $SU(3)$ symmetric breaking hyperfine splittings, respectively. 
By fitting these LECs, one can use the theory, for example, to compute \\
(a) hyperfine splittings,
\begin{align}\label{sss}
&m_{A^*_q}-m_{A_q}=\Delta_A+\Delta_A^{(a)}m_q+\Delta_A^{(\sigma)}\overline{m}+\Sigma_{A^*_q}-\Sigma_{A_q};
\end{align}
(b) $SU(3)$ flavor splittings,
\begin{align}
&m_{A_s}-m_{A_n}=a_A(m_s-m_n)-\frac{3}{4}\Delta_A^{(a)}(m_s-m_n)+\Sigma_{A_s}-\Sigma_{A_n};
\end{align}
(c) spin-average masses,
\begin{align}
&(m_{A_q}+3m_{A^*_q})/4=\delta_A+a_A m_q+\sigma_A \overline{m}+(\Sigma_{A_q}+3\Sigma_{A^*_q})/4;
\end{align}
(d) $SU(3)$-violating hyperfine splittings, 
\begin{align}\label{su3violate}
(m_{A^*_s}-m_{A_s})-(m_{A^*_n}-m_{A_n})&=\Delta_A^{(a)} (m_s-m_n)+(\Sigma_{A^*_s}-\Sigma_{A_s})-(\Sigma_{A^*_n}-\Sigma_{A_n});
\end{align}
(e) spin-average strange and nonstrange mass differences,
\begin{align}\label{bbb}
(m_{A_s}+3m_{A^*_s})/4-(m_{A_n}+3m_{A^*_n})/4&=a_A (m_s-m_n)+(\Sigma_{A_s}-\Sigma_{A_n}+3\Sigma_{A^*_s}-3\Sigma_{A^*_n})/4,
\end{align}
in the odd- and even-parity charmed meson sectors. 
It can also be used to predict the analog quantities in the bottom meson sector. This requires rescaling
hyperfine operators by  the mass ratio of charm and bottom quarks, $m_c/m_b$; see Ref. \cite{Alhakami} for details. 
\section{Results and conclusion}
There are 12 unknown LECs in Eq.~\eqref{eq1} describing eight charmed meson masses in the isospin limit.
It is, thus, hard to fix them using available data alone. 
To overcome this, LECs can be grouped into the following linear combinations \cite{Alhakami},
\begin{align}\label{m1}
&\eta_A=\delta_A+(\frac{a_A}{3}+\sigma_A)\, \overline{m},~~\xi_A=\Delta_A+(\frac{\Delta^{(a)}_A}{3}+\Delta^{(\sigma)}_A)\, \overline{m},\\[2ex]\label{m2}
&L_A=(m_s-m_n)\,a_A,~~~~~~~T_A=(m_s-m_n)\,\Delta^{(a)}_A,
\end{align}
where terms in $\eta_A$ and $\xi_A$ ($L_A$ and $T_A$) preserve (violate) $SU(3)$ flavor symmetry.
The combinations $\xi_A$ and $T_A$ contain heavy quark spin-symmetry-violating
operators. In terms of these combinations, 
Eq.~\eqref{eq1} can be written as 
\begin{equation}\label{D}
m_{A^{(*)}_q}=\eta_A+\frac{d^{(*)}}{4}\xi_A+\frac{\alpha_q}{3} L_A+\frac{\beta^{(*)}_q}{2} T_A+\Sigma_{A^{(*)}_q},
\end{equation}
where $\alpha_q$ and $\beta^{(*)}_q$ are $\alpha_n=-1$, $\alpha_s=2$, $\beta_n=1/2$, $\beta_s=-1$, 
$\beta^*_n=-1/6$, and $\beta^*_s=1/3$. 

\begin{figure}[h!]
\centering
\includegraphics[scale=0.3,width=0.7\textwidth]{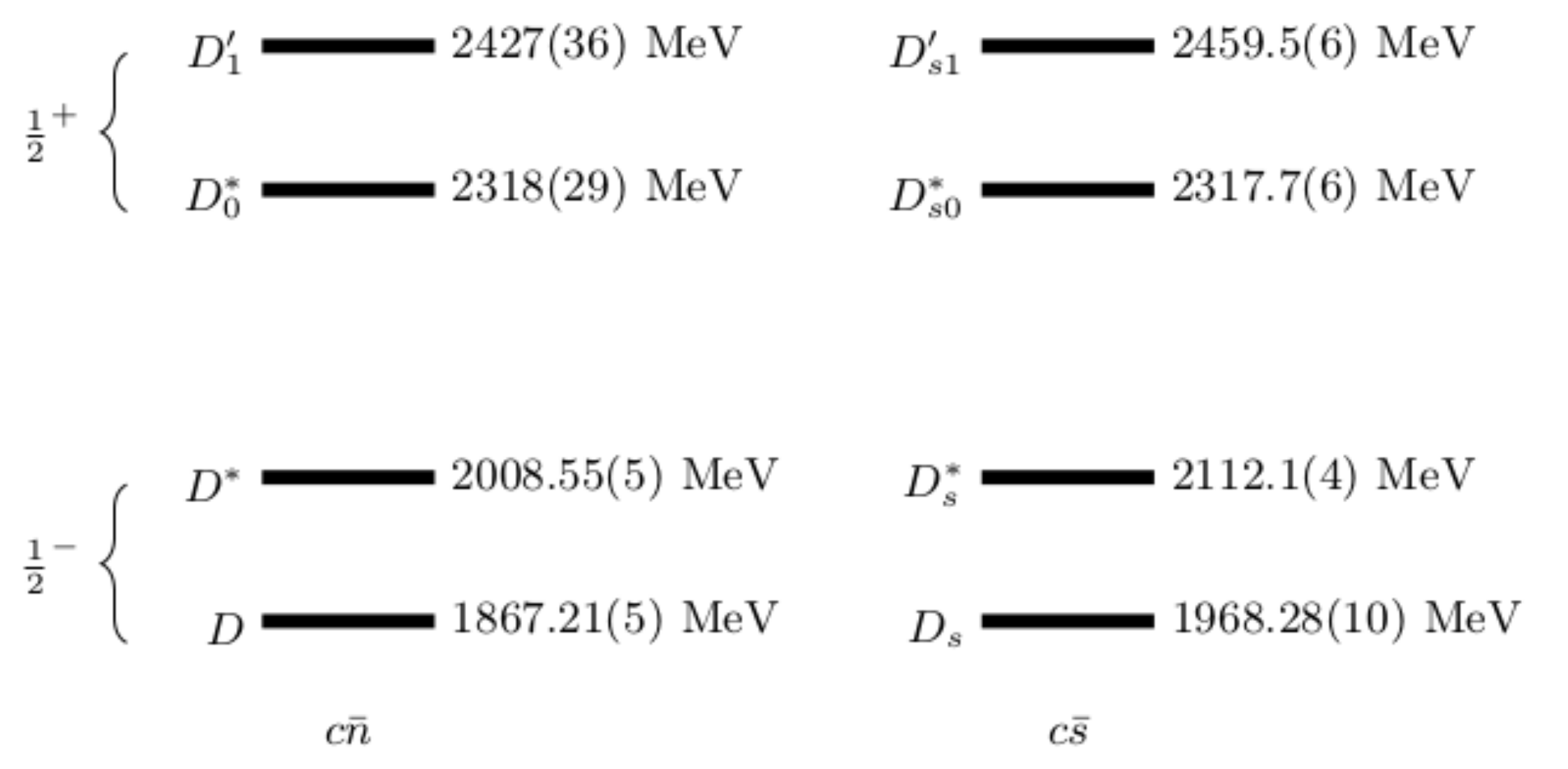}
\caption{The representation of the masses of the charmed meson states of $\frac{1}{2}^-$  and
$\frac{1}{2}^+$ doublets. All masses are taken from the PDG \cite{pdg12} excluding the mass of $D^\prime_1$, which is reported by the BELLE 
Collaboration \cite{19}. We only take the isospin average of $D^0$ and $D^\pm$ ($D^{*0}$ and $D^{*\pm}$)
to obtain the mass of nonstrange ground state $D$ ($D^*$); for details please refer to the text.}
\label{M11parity}
\end{figure}

Now, the number of unknown coefficients in  Eq.~\eqref{D} is 8, which equals the number of 
the observed charmed mesons shown in Fig.~\ref{M11parity}.                    
By using physical values in evaluating chiral loop functions in Eq.~\eqref{D}, as done in Ref. \cite{Alhakami}, one can extract the unique values
for the parameters given in Eqs.~\eqref{m1} and \eqref{m2}. 

It is clear from Eq.~\eqref{m2} that the available experimental information is enough to fix the LECs
$a_H$, $\Delta^{(a)}_H$ and $a_S$, $\Delta^{(a)}_S$
of the both odd- and even-parity sectors. 
Nature, however, cannot help us disentangle chirally symmetric coefficients $\delta_A$, $\Delta_A$ in Eq.~\eqref{m1} from chiral
breaking terms, more precisely $\sigma_A$ and $\Delta^{(\sigma)}_A$  as $a_A$ and $\Delta^{(a)}_A$ already fixed by experiment. 
To make further separations of the LECs in Eq.~\eqref{m1} in the odd- and even-parity sectors, 
lattice calculations on charmed mesons ground and excited states with different quark masses are required.
We will show below how to use experimental and lattice data on the charmed meson masses to fit the LECs  that appear in the mass expansion given in Eq.~\eqref{eq1}.

Let us first use the experimental information to extract the parameters given in Eqs.~\eqref{m1} and \eqref{m2}. In our fit, the empirical values we use  are two masses of the ground-state nonstrange mesons in the isospin limit, 
two masses of the excited neutral charmed mesons, 
which are chosen due to their relatively small errors in comparison with the excited charged counterpart,
and four masses of strange mesons
from both sectors; see Fig.~\ref{M11parity}.  
In our calculations, the following physical values are used: $m_n=4$ MeV, $m_s=130$ MeV, $m_{\pi}=140$ MeV, $m_K = 495$ MeV, $m_\eta = 547$ MeV, 
and $f=92.4$ MeV \cite{pdg12}. 
For coupling constants, we use the experimental determined values 
$g=0.64\pm0.075$ and $h=0.56\pm0.04$ \cite{cdgn12}.
The coupling constant $g^\prime$ is experimentally unknown, and the computed LQCD value $g^\prime=-0.122(8)(6)$ \cite{glattice}
is used in this work.
In our previous work \cite{Alhakami}, the normalization scale was set to 
the average of pion and kaon masses, $\mu=317$ MeV.
It is worth mentioning that in our approach the extracted parameters and quantities derived from them, e.g., mass splittings, are smoothly varying 
with the $\mu$-scale and their numerical values are in agreement within the associated uncertainties. Therefore, performing calculations at any other values 
of the $\mu$-scale will not make much difference. Here, we will use $\mu=1$ GeV.

To fit parameters in Eq.~\eqref{D} to the experiment, we need to define the experimental residual masses.
For this, we choose $m_D$, the mass of pseudoscalar nonstrange charmed meson,
as the reference mass, which yields the following values for charmed meson residual masses:
\begin{equation}\label{rm1}
\begin{split}
&m_{H_n}=0(0)~ \text{MeV}, ~~~~~~~~~m_{H_s}=101.1(1) ~\text{MeV},\\
&m_{H^*_n}=141.3(7) ~\text{MeV}, ~~~~m_{H^*_s}= 244.9(4)~\text{MeV},\\
&m_{S_n}=451(29)~ \text{MeV}, ~~~~~m_{S_s}= 450.5(6)~ \text{MeV},\\
&m_{S^*_n}=560(36) ~\text{MeV},~~~~~ m_{S^*_s}=592.3(6)~\text{MeV}.
\end{split}
\end{equation}
Using physical values of charmed meson masses,
pseudo-Goldstone boson masses, and coupling constants in chiral loop functions, 
one gets 
\begin{align}\label{dh}
&\eta_H=228(46)~ \text{MeV}, ~~~~~\xi_H=88(20)~ \text{MeV},\\\label{lh}
&L_H=262(28)~ \text{MeV},  ~~~~T_H=-138 (41)~ \text{MeV},\\\label{ds}
&\eta_S=542(20)~ \text{MeV},  ~~~~~\xi_S=110(33)~ \text{MeV},\\\label{ls}
&L_S=-42(31)~ \text{MeV},  ~~~~T_S=42(49)~ \text{MeV},
\end{align}
from fitting the residual mass expression in Eq.~\eqref{D} to the corresponding experimental masses  in Eq.~\eqref{rm1}. 
The  associated uncertainties with the fitted parameters, which include the experimental errors of charmed meson masses and 
coupling constants and the error on the coupling $g^\prime$ from LQCD,
are dominated by the uncertainty in the $0^+$ and $1^+$ nonstrange masses.
Therefore, improved experiments on these mesons are needed to reduce the errors.

From the above extracted values of $L$'s and $T$'s, see Eqs.~\eqref{lh} and \eqref{ls}, one can fix the following LECs, see Eq.\eqref{m2}:
\begin{equation}
\begin{split}\label{l1}
&a_H=2.08(22),~~~~~\Delta^{(a)}_H=-1.10(33),~~~~~ a_S=-0.33(25),~~~~\Delta^{(a)}_S=0.33(39).
\end{split}
\end{equation}
To extract the other LECs, we will use lattice calculations on charmed meson spectroscopy undertaken in Ref. \cite{lattice}.
There, the computations were performed using three different lattice spacings and several light quark masses.
In this paper, we use the values extracted in ensemble D defined in Ref. \cite{lattice} that have the lightest pion masses ($m_\pi\lesssim$ 250 MeV
that lies within the range of validity of $\chi$PT)
in our fit of LECs. In Table \ref{table:lattice}, we present the continuum masses of odd- and even-parity charmed mesons computed
at nonphysical pion masses. The shown values are obtained by performing a continuum extrapolation 
at the relevant nonphysical pion masses using strategy 3 illustrated there \cite{st3}.
For the nonstrange ground-state charmed meson, the authors of Ref. \cite{lattice} used its mass as an input to fix the charm quark mass for each ensemble, so
in our fit, we will use the experimental value shown in Fig.~\ref{M11parity}. 
In their work, strange valence quark mass was chosen to be close to its physical value. This was achieved by 
reproducing the physical value of  $2m^2_K-m^2_{\pi}$ using measured pion and kaon masses in each ensemble.
In leading order chiral perturbation theory, this quantity represents the strange light quark mass and is insensitive 
to the mass of nonstrange light quark flavor. 
Consequently, one can use the computed values of pion mass in ensemble D
to extract the corresponding masses of kaon and eta particles. 
This is simply done by using the mass relations
$((2m^2_K-m^2_{\pi})_{\text{phys}}+m^2_{\pi,L})/2$ and 
$(2(2m^2_K-m^2_{\pi})_{\text{phys}}+m^2_{\pi,L})/3$ to get
$m^2_K$ and $m^2_\eta$, respectively, where $m_{\pi,L}$ is the lattice measured pion mass; see Table \ref{table:lattice}.
The uncertainties associated with the lattice determination of these masses are negligible at our level of precision.
\begin{table}[h!]
\def\arraystretch{1.5}
\begin{tabular}[t]{|c|cccccccccccc|}
\toprule
\hline
Ensemble& $m_{D^*}$&$m_{D_s}$& $m_{D^*_s}$&$m_{D^*_0}$&$m_{D^\prime_1}$&$m_{D^*_{s0}}$& $m_{D^\prime_{s1}}$& $m_{\pi}$&$m_K$&$m_{\eta}$&$\overline{m}$&$m_s-m_n$\\
\hline
\bottomrule
D15.48 & 2029.0(7.0)& 1962.6(2.8)& 2119.3(3.8) & 2351(10) & 2490(15) & 2400(11) & 2565(10)& 224 & 513 & 579 & 392 & 377\\
D20.48 & 2030.0(7.1)&1959.9(2.8)&2117.7(3.9)&2364(10) & 2503(15) & 2404(11) & 2570(10)& 257 & 521 & 583 & 395 & 376 \\
\hline
\end{tabular}
\caption{The listed numerical values are in MeV. The charmed meson masses are obtained using strategy 3 \cite{st3,lattice}. 
The nonstrange ground-state charmed meson mass, $m_D$, was used in Ref. \cite{lattice} to tune the charm quark mass in their lattice computations.
In our calculation, we use the experimental value shown in Fig.~\ref{M11parity} for this nonmeasured lattice mass.}
\label{table:lattice}
\end{table}

\begin{figure}[h!]
\subfloat[ ]{\includegraphics[width = 3in]{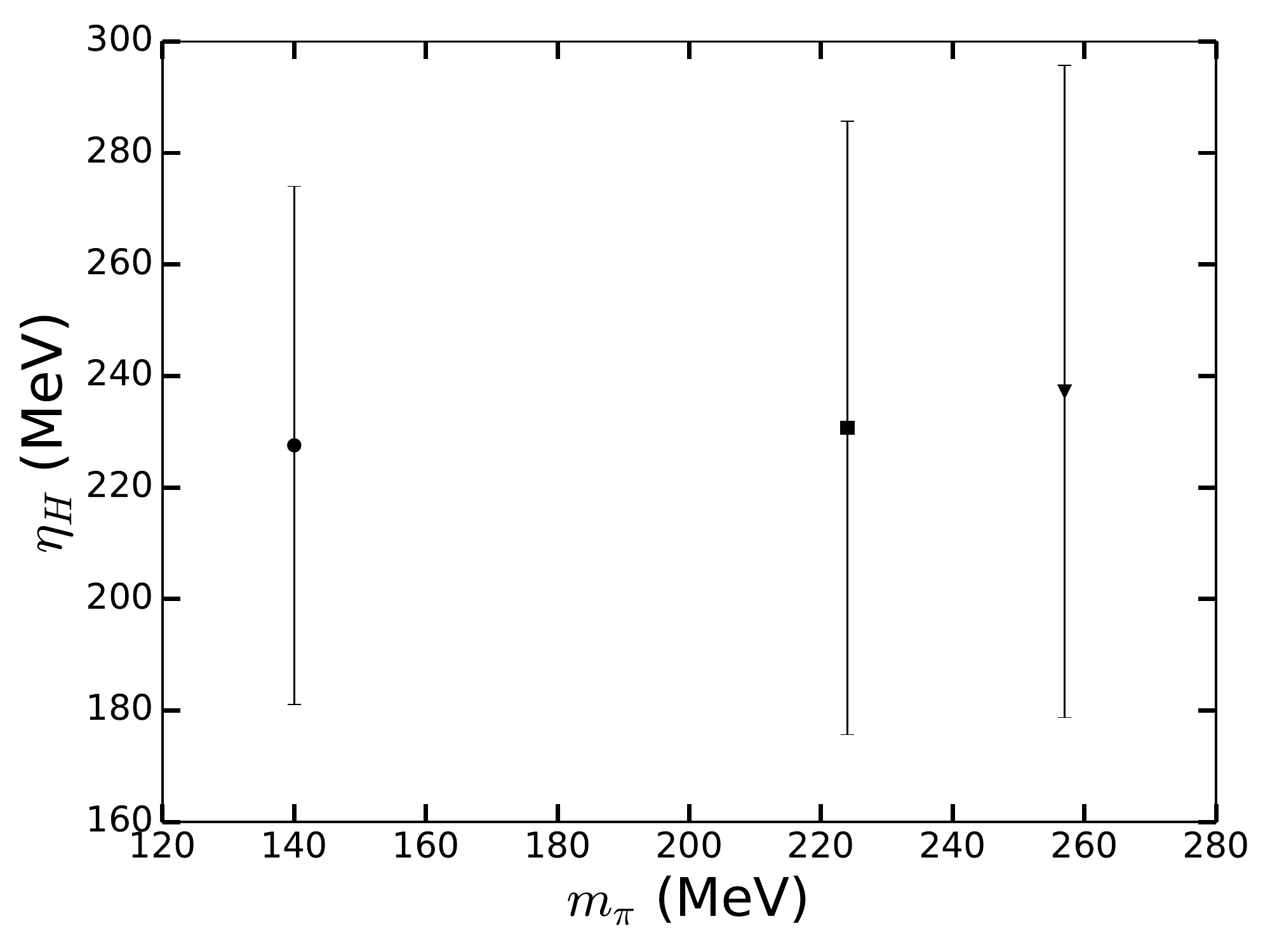}} 
\subfloat[ ]{\includegraphics[width = 3in]{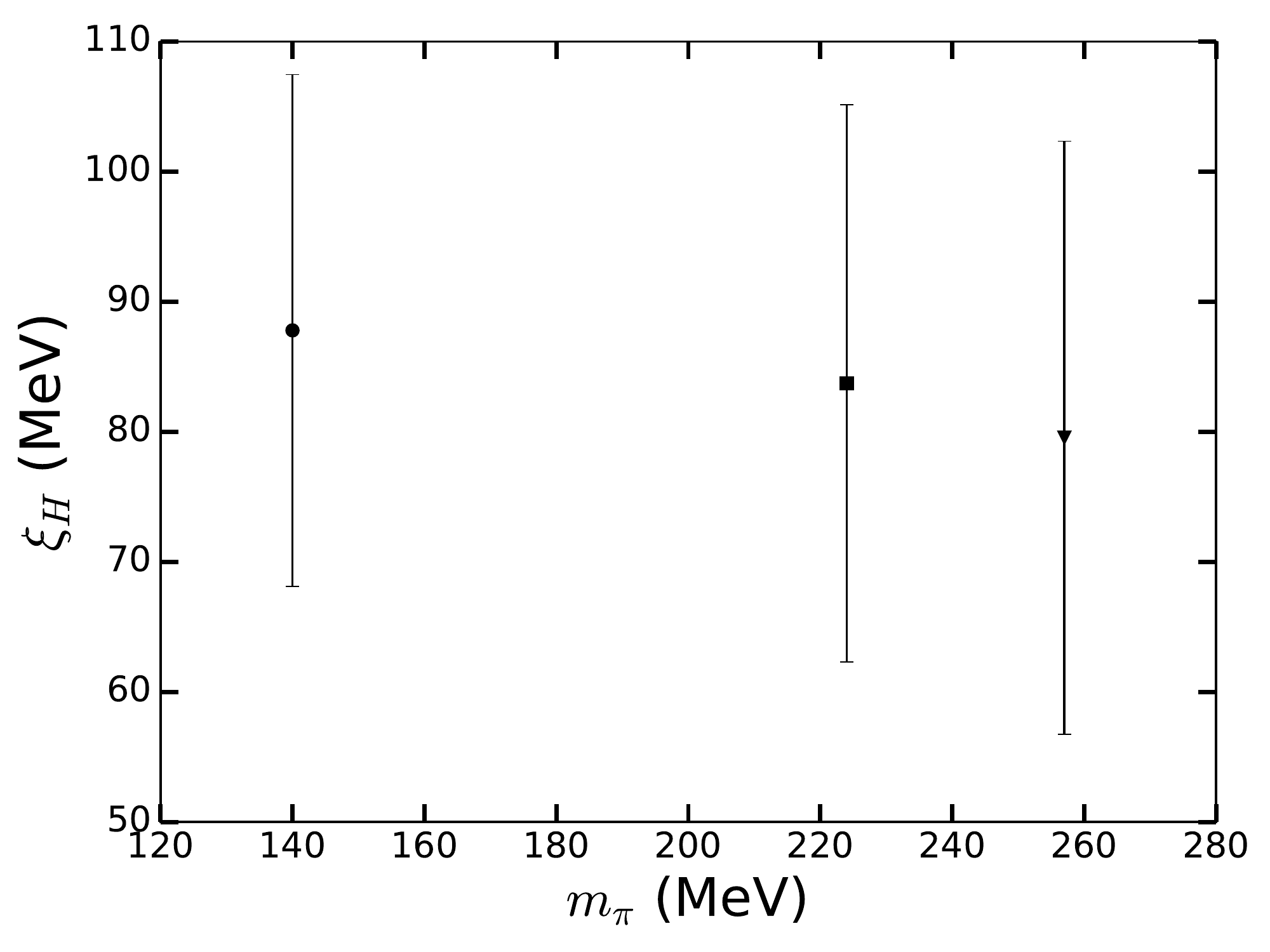}}\\
\subfloat[ ]{\includegraphics[width = 3in]{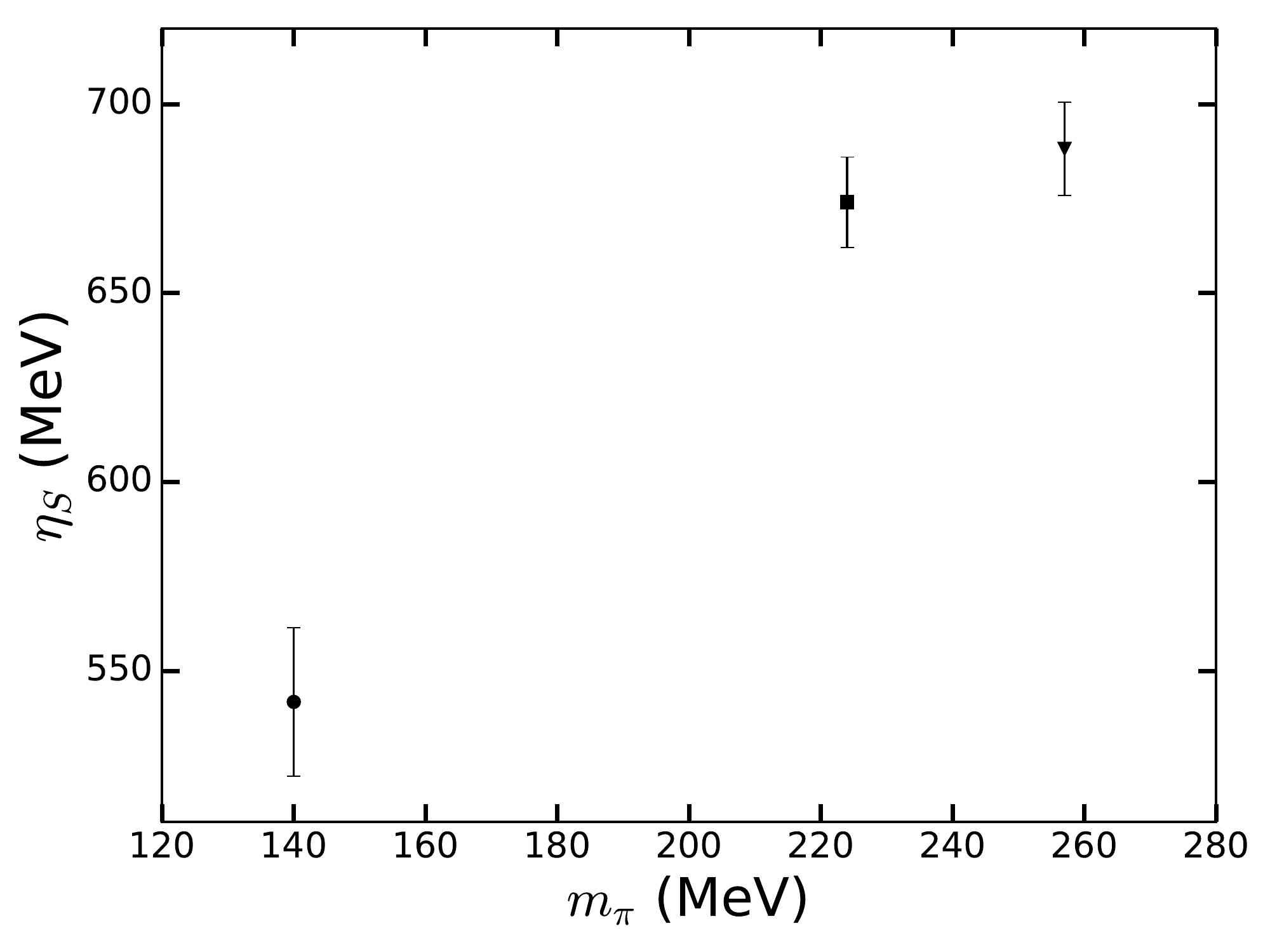}}
\subfloat[ ]{\includegraphics[width = 3in]{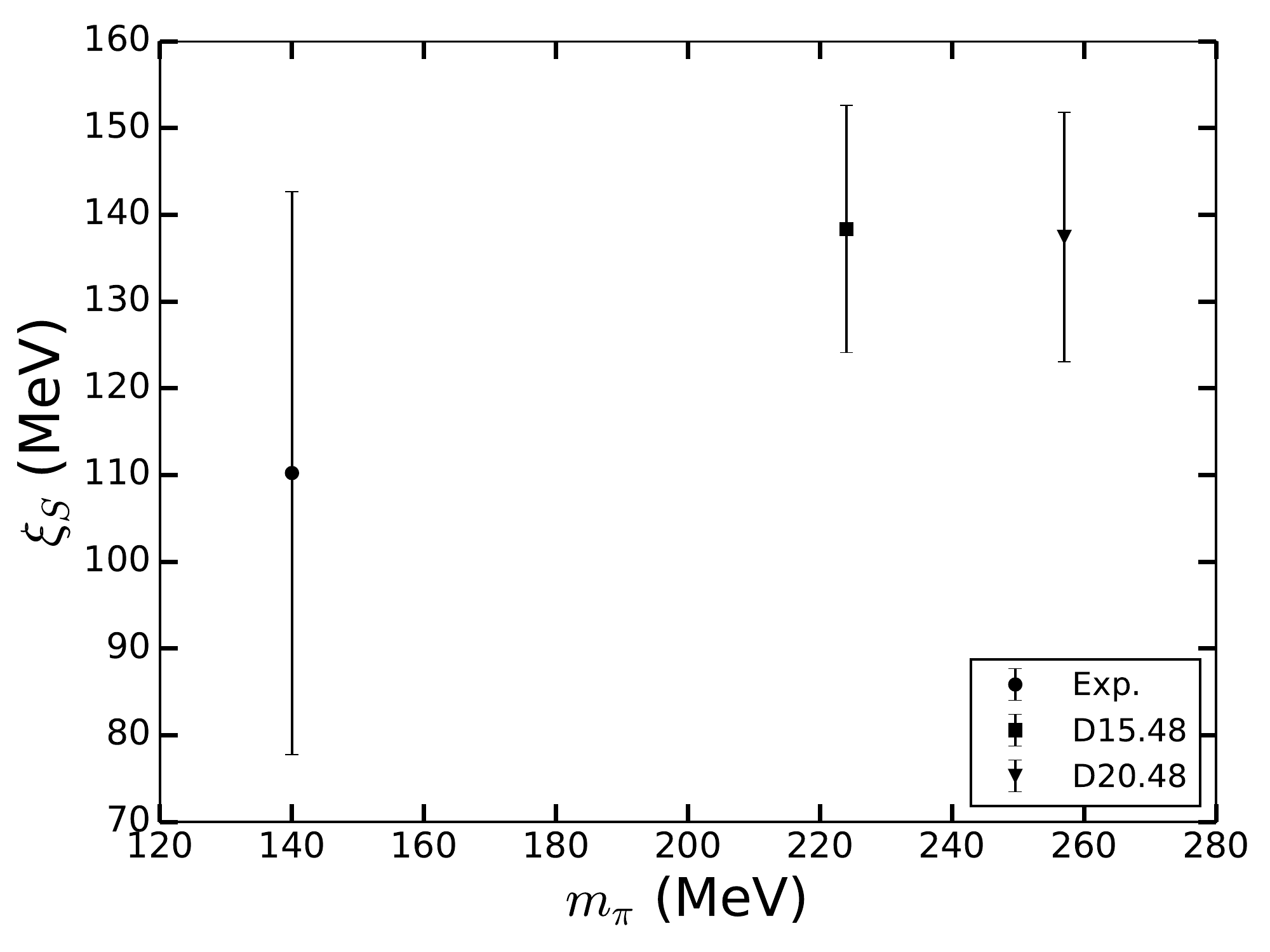}} 
\caption{Extracted numerical values for the combinations (a) $\eta_H$, (b) $\xi_H$, (c) $\eta_S$, and (d) $\xi_S$
are plotted against the corresponding pion masses.
    Different symbols are given to the experiment and two 
    lattice ensembles according to the key in the $\xi_S$ plot.}
\label{fig7}
\end{figure}

Using lattice data from Table \ref{table:lattice}, extracted values of parameters given in Eq.~\eqref{m1}
are shown in Figs.~\ref{fig7}(a)--\ref{fig7}(d) together with that obtained using experimental values; see Eqs.~\eqref{dh} and \eqref{ds}. 
To fit these parameters, a constrained fitting procedure \cite{prior} is employed with priors on the LECs constructing them. 
For LECs  $a_A$ and  $\Delta^{(a)}_A$, their extracted values in Eq.~\eqref{l1} are used as priors information.
On the other hand, the charmed meson spectrum constraints the combinations of the other unphysical LECs, i.e., $\delta_A$, $\Delta_A$, $\sigma_A$, and $ \Delta^{(\sigma)}_A$; 
therefore, it is appropriate to use broad priors for them.
We set $0\pm 1000$ MeV ($0\pm 1000$) as priors on 
 $\delta_A$ and $\Delta_A$ ($\sigma_A$ and $ \Delta^{(\sigma)}_A$).
Performing a least chi-squared fit to these parameters yields
\begin{equation}
\begin{split}\label{l3}
&\delta_H=224(74)~\text{MeV},~~~ \sigma_H=-0.67(25)~~~~\Delta_H=91(32)~\text{MeV},~~~~\Delta^{(\sigma)}_H=0.34(15),\\
&\delta_S=466(31)~\text{MeV},~~~~\sigma_S=0.66(12)~~~~~~~\Delta_S=95(50)~\text{MeV},~~~~\Delta^{(\sigma)}_S=-0.003(184),
\end{split}
\end{equation}
where associated uncertainties include the experimental errors of charmed meson masses and coupling constants and errors from 
lattice data on charmed meson masses. 

The extracted values given in Eqs.~\eqref{l1} and \eqref{l3} are consistent with the perturbative expansion of the theory.
They yield the following values for the
residual masses, 
\begin{equation}
\begin{split}\label{extrp}
&m_{H_n}=-1(67)~\text{MeV},~~~~~~m_{H_s}=101(58)~\text{MeV},\\&m_{H^*_n}=141(105)~\text{MeV},~~~~m_{H^*_s}=245(68)~\text{MeV},\\
&m_{S_n}=451(37)~\text{MeV}, ~~~~~~m_{S_s}= 451(27)~\text{MeV},\\&m_{S^*_n}=560(46)~\text{MeV}, ~~~~~~m_{S^*_s}=593(32)~\text{MeV},
\end{split}
\end{equation}
which are compatible with the experimental values given in Eq.~\eqref{rm1}.
To shrink the uncertainties on the determined LECs [Eqs.~\eqref{l1} and \eqref{l3}] and, hence, the extrapolated residual masses [Eq.~\eqref{extrp}],
accurate experimental and lattice results on charmed meson masses are needed.  

By fitting LECs of the effective Lagrangian, we increased the usefulness of HM$\chi$PT to other applications of heavy-light meson systems,
e.g., calculating masses and strong mass splittings that are shown in Eqs.~\eqref{sss}-\eqref{bbb} for 
the lowest-lying $S$- and $P$-wave states of charmed and bottom mesons. 
\begin{acknowledgments}

It is a pleasure to thank Krzysztof Cichy, Chris Bouchard, and  Javad Komijani for very helpful discussions.

\end{acknowledgments}


\begin{thebibliography}
\frenchspacing
\bibitem{wise}  M. Wise, Phys. Rev. D \textbf{45}, R2188 (1992).
\bibitem{4} E. Jenkins, Nucl. Phys. \textbf{B412}, 181 (1994).
\bibitem{PCh} P. L. Cho, Nucl. Phys. \textbf{B396}, 183 (1993); \textbf{B421}, 683(E) (1994).
\bibitem{FalkLuke} A. F. Falk and M. E. Luke, Phys. Lett. B \textbf{292}, 119 (1992).
\bibitem{cas97} R. Casalbuoni \textit{et al.}, Phys. Rep. \textbf{281}, 145 (1997).
\bibitem{ms05} T. Mehen and R. Springer, Phys. Rev. D \textbf{72}, 034006 (2005).
\bibitem{absu07} B. Ananthanarayana, S. Banerjee, K. Shivaraja, and A.Upadhyaya, Phys. Lett. B \textbf{651}, 124 (2007).
\bibitem{Alhakami} M. H. Alhakami, Phys. Rev. D \textbf{93}, 094007 (2016).
\bibitem{lattice} K. Cichy, M. Kalinowski, and M. Wagner, Phys. Rev. D \textbf{94}, 094503 (2016).
\bibitem{rm} The residual mass is taken to be the difference between the experimental mass and an arbitrarily chosen reference mass of $O(m_Q)$ \cite{ms05}. 
\bibitem{Ffunction} S. Scherer, Adv. Nucl. Phys. \textbf{27}, 277 (2003). 
\bibitem{pdg12} C. Patrignani \textit{et al}. (Particle Data Group Collaboration), Chin. Phys. C, \textbf{40}, 100001 (2016) and 2017 update, \url{http://pdg.lbl.gov/}. 
\bibitem{19} K. Abe \textit{et al.} (BELLE Collaboration), Phys. Rev. D \textbf{69}, 112002 (2004).
\bibitem{cdgn12} P. Colangelo, F. De Fazio, F. Giannuzzi, and S. Nicotri, Phys. Rev. D \textbf{86}, 054024 (2012).
\bibitem{glattice} B. Blossier, N. Garron, and A. Gerardin, Eur. Phys. J. C \textbf{75}, 103 (2015). 
\bibitem{st3}
In Ref. \cite{lattice}, strategy 3 gives us the continuum charmed meson mass at nonphysical pion mass using the relation
$m_D(m_\pi)=m^c_D+\alpha (m^2_\pi-m^2_{\pi,\mathrm{exp}})$ where $m^c_D$ is the continuum charmed meson mass at 
physical mass of the neutral pion $m_{\pi,\mathrm{exp}}$, which is 135 MeV, and $\alpha$ is a fit parameter defined there. 
In Table \ref{table:lattice} of the present paper, the continuum charmed meson masses at nonphysical pion masses in each ensemble are calculated using this relation by taking the
GeV value of the corresponding $m^c_D$ 
and  the relevant value of coefficient $\alpha$ 
from Tables III and IV of
Ref. \cite{lattice}. 
\bibitem{prior} G. P. Lepage, lsqfit v4.8.5.1, \url{https://doi.org/10.5281/zenodo.10236}.
\end{thebibliography}
\end{document}